\documentclass[aps,twocolumn,superscriptaddress,prd]{revtex4-2}

\usepackage{bm,graphicx,textcomp,amssymb,amsmath,dcolumn,xcolor}
\usepackage{physics}
\usepackage{subfigure}

\newcommand{\bea}{\begin{eqnarray}}
\newcommand{\ea}{\end{eqnarray}}
\newcommand{\eea}{\end{eqnarray}}

\usepackage{hyperref} 
\hypersetup{ 
    colorlinks=true,
    citecolor=blue,
	linkcolor=blue
}

\begin{document}

\title{Disorder-induced localisation in the Mott-Hubbard model}

\author{R.~Kristers Knip\v{s}is}
\affiliation{Helmholtz-Zentrum Dresden-Rossendorf, 
Bautzner Landstra{\ss}e 400, 01328 Dresden, Germany}
\affiliation{Institut f\"ur Theoretische Physik, 
Technische Universit\"at Dresden, 01062 Dresden, Germany}

\author{F.~Queisser}
\affiliation{Helmholtz-Zentrum Dresden-Rossendorf, 
Bautzner Landstra{\ss}e 400, 01328 Dresden, Germany}
\affiliation{Institut f\"ur Theoretische Physik, 
Technische Universit\"at Dresden, 01062 Dresden, Germany}

\author{J.~Jayabalan}
\affiliation{Fakultät für Physik und Center for Nanointegration
(CENIDE), Universität Duisburg-Essen, Lotharstr.~1, 47057
Duisburg, Germany}

\author{G.~Reecht}
\affiliation{Fakultät für Physik und Center for Nanointegration
(CENIDE), Universität Duisburg-Essen, Lotharstr.~1, 47057
Duisburg, Germany}








\author{M.~Gruber}
\affiliation{Fakultät für Physik und Center for Nanointegration
(CENIDE), Universität Duisburg-Essen, Lotharstr.~1, 47057
Duisburg, Germany}

\author{U.~Bovensiepen}
\affiliation{Fakultät für Physik und Center for Nanointegration
(CENIDE), Universität Duisburg-Essen, Lotharstr.~1, 47057
Duisburg, Germany}

\author{R.~Sch\"utzhold}
\affiliation{Helmholtz-Zentrum Dresden-Rossendorf, 
Bautzner Landstra{\ss}e 400, 01328 Dresden, Germany}
\affiliation{Institut f\"ur Theoretische Physik, 
Technische Universit\"at Dresden, 01062 Dresden, Germany}

\date{\today}

\begin{abstract}
For the 
Fermi-Hubbard model in the Mott insulator phase, we employ the hierarchy of correlations 
to study how doublon and holon quasi-particle excitations are affected by adding disorder 
to the system. 
We study two types of disorder: charge disorder, in the form of 
on-site potential randomness; and spin disorder, in the form of a fixed, 
randomly generated background spin arrangement. 
By analysing the quasi-particle eigen-spectra and quantifying the degree to which the 
corresponding eigen-states localise, we find both an energetic and spatial separation 
between localised and delocalised states in the charge disorder. 
In contrast, the spin disorder results in localised states throughout the quasi-particle bands. 
Finally, we repeat our calculations using strong-coupling perturbation theory, 
and compare the results obtained from both methods.
\end{abstract}

\maketitle


\section{Introduction}

Going from textbooks to real life, one quickly finds the perfectly ordered systems 
of the former affected to varying degrees of imperfections and disorder in the latter. 
This has historically been an unwanted effect to be eliminated or worked around. 
In recent decades, however, as the interest in 
suppressing thermalisation in quantum systems has grown, 
considerable 
attention has been paid to disordered systems.

In Anderson's seminal paper \cite{anderson1958absence}, it was argued that an on-site 
disorder can localise non-interacting particles in a lattice, eliminating long-range 
transport in the system at zero temperature. 
Subsequently, the same conclusion was extended to weakly interacting systems \cite{fleishman1980interactions}. 
Since then, localisation has been observed in various photonic 
\cite{chabanov2000statistical, lahini2008anderson, schwartz2007transport}, 
acoustic \cite{weaver1990anderson, hu2008localization}, 
and matter systems with suppressed inter-particle interactions 
\cite{kondov2011three, billy2008direct}.

On the other hand, it was long unclear whether localisation would survive in a 
strongly interacting system at nonzero temperatures. 
An answer to the affirmative has been suggested in the form of many-body localisation (MBL), 
where the existence of quasi-local integrals of motion point towards ergodicity breaking 
even at finite temperatures 
\cite{gornyi2005interacting, basko2006metal, abanin2019colloquium, roy2025fock}. 
This remarkable theory quickly gained experimental support \cite{schreiber2015observation}. 
On the other hand, coupling to phonon baths is known to destroy MBL, 
and it is also disputed whether the MBL phase is stable in more than one dimension.

Motivated by these questions, 
our starting point is the Fermi-Hubbard model \cite{hubbard1963electron} 
in a 2D lattice 
\begin{equation}
    \hat{H} = -\frac{1}{Z}\sum_{\mu, \nu, s} T_{\mu\nu} \hat{c}_{\mu,s}^\dagger \hat{c}_{\nu,s} + U\sum_{\mu}\hat{n}_{\mu,\uparrow}\hat{n}_{\mu, \downarrow},
\end{equation}

\noindent
where $\mu, \nu$ are site indices, $T_{\mu\nu} = T$ for nearest neighbour sites and 
zero otherwise, $Z$ is the coordination number of the lattice, 
$\hat{c}_{\mu,s}^\dagger$ ($\hat{c}_{\mu,s}$) is the creation (annihilation) 
operator for an electron of spin $s$ at site $\mu$, and $U$ is the on-site interaction energy. 

\section{Setup}

We consider two-dimensional lattices with different coordination numbers, namely, 
the hexagonal ($Z=6$), square ($Z=4$), and honeycomb ($Z=3$) lattices. 
Throughout this paper, we enforce periodic boundary conditions, 
and only consider lattices with equal sizes in both dimensions, $N = N_x = N_y$, 
and, unless stated otherwise, we take $N=25$. 
Note that $\hbar = 1$ throughout this paper.

We consider the limit of weak hopping strength $T \ll U$, such that also $T \gg T^2 / U$. 
We restrict our considerations to timescales $\sim 1/T$, and in particular such that we do 
not reach 
the  spin dynamics timescale $\sim U/T^2$. 
This ensures that the spin 
backgrounds considered in this paper can be well approximated to be static.

In this paper, we consider two types of disorder, namely, charge disorder and spin disorder. 
The former is defined by a spin-independent on-site potential,
\begin{equation}
    \hat{H}_{c} = \sum_{\mu} V_{\mu} \left( \hat{n}_{\mu, \uparrow} + \hat{n}_{\mu, \downarrow} \right), \label{eq:charge_disorder_hamiltonian}
\end{equation}
where $V_{\mu}$ is assigned with probability $r$ to each lattice site to be either $0$ or $V$. 
This is in contrast to typical Anderson localisation where $V_{\mu}$ is sampled from a 
continuous distribution.

Meanwhile, spin disorder is defined by an on-site potential that introduces a spin-splitting,
\begin{equation}
    \hat{H}_{s} = \sum_{\mu} v_{\mu} \left( \hat{n}_{\mu, \uparrow} - \hat{n}_{\mu, \downarrow} \right), \label{eq:spin_splitting_hamiltonian}
\end{equation}
where $v_{\mu} \in \{-v, v\}$ is again randomly chosen at each site, thereby energetically 
favouring a particular spin-orientation. We choose $T \gg v \gg T^2/U$, 
such that spin 
dynamics are frozen in the timescales considered in this paper. 
In practice, to make calculations tractable, the spin-splitting is imposed implicitly by 
only considering states in the lowest energy subspace. 
Consequently, it should be kept in mind that the corresponding results only hold on 
sufficiently short timescales.

\section{Hierarchy of correlations}

\subsection{Factorised equations of motion}

The hierarchy of correlations is an expansion in orders of $1/Z$ where $Z\gg 1 $ is the 
coordination number of the lattice \cite{navez2010emergence}. 
In this paper we are interested in one-point density matrices 
and two-point correlators, in other words, 
the objects of interest are the reduced density matrices $\hat{\rho}_{\mu}$ 
and $\hat{\rho}_{\mu\nu}$. 
In fact, a hierarchy in orders of $1/Z$ requires considering only the correlated parts 
of the density matrices. 
For instance, the two-point object of interest is 
$\hat{\rho}_{\mu\nu}^{corr} = \hat{\rho}_{\mu\nu} - \hat{\rho}_{\mu}\hat{\rho}_{\nu}$, 
the three-point reduced density matrix of interest would be 
$\hat{\rho}_{\mu\nu\lambda}^{corr} = \hat{\rho}_{\mu\nu\lambda} - \hat{\rho}_{\mu}\hat{\rho}_{\nu}\hat{\rho}_{\lambda} - \hat{\rho}_{\mu\nu}^{corr}\hat{\rho}_{\lambda} - \hat{\rho}_{\mu\lambda}^{corr}\hat{\rho}_{\nu} - \hat{\rho}_{\nu\lambda}^{corr}\hat{\rho}_{\mu}$, and so on.

Then, the time evolution of the relevant density matrices follows the hierarchy \cite{navez2010emergence}
\begin{align}
    i\partial_t \hat{\rho}_{\mu} &= f_1(\hat{\rho}_{\mu}, \hat{\rho}_{\mu\nu}^{corr}) \sim \mathcal{O}(1),\\
    i\partial_t \hat{\rho}_{\mu\nu}^{corr} &= f_2(\hat{\rho}_{\mu}, \hat{\rho}_{\mu\nu}^{corr}, \hat{\rho}_{\mu\nu\lambda}^{corr}) \sim \mathcal{O}(1/Z).
\end{align}

In fact, as long as the density matrices also meet corresponding initial conditions $\hat{\rho}_{\mu} \sim \mathcal{O}(1)$, $\hat{\rho}_{\mu\nu}^{corr} \sim \mathcal{O}(1/Z)$ 
and so on, this hierarchy is also conserved at subsequent times limited only by the growth 
of unstable modes (if present) \cite{navez2010emergence}. 
This observation allows us to truncate to a particular order in $1/Z$. 
In this paper, we truncate to first order $1/Z$, meaning that the contributions of all 
density matrices beyond two-point can be neglected.

At this point, we can specify the correlators of interest. 
When considering quasi-particles on a half-filled background, the relevant operators 
are the dressed fermionic operators

\begin{equation}
    \hat{C}_{\mu, s}^I = \hat{N}_{\mu,\bar{s}}^I \hat{c}_{\mu, s} =
    \begin{cases}
        \hat{n}_{\mu, \bar{s}}\hat{c}_{\mu, s} &I=1,\\
        (1-\hat{n}_{\mu, \bar{s}})\hat{c}_{\mu, s} &I=0,\\
    \end{cases}
\end{equation}

\noindent
which act as precursors for holon ($I=0$) and doublon ($I=1$) quasiparticles.

We now consider the two-point correlators 
$\langle \hat{C}_{\mu, s}^{I \dagger} \hat{C}_{\nu, s}^J\rangle^{\mathrm{corr}}$. 
In fact, to first order $1/Z$, we can equivalently employ the factorisation 
ansatz \cite{navez2014quasi} to write
\begin{equation}
    \langle \hat{C}_{\mu, s}^{I \dagger} \hat{C}_{\nu, s}^J\rangle^{corr} = \left(p_{\mu, s}^{I}\right)^* p_{\nu, s}^{J} \label{eq:factorisation_ansatz}
\end{equation}

\noindent
such that to get the time dependence of  
$\langle \hat{C}_{\mu, s}^{I \dagger} \hat{C}_{\nu, s}^J\rangle^{corr}$ 
we need only to solve the much simpler set of equations

\begin{equation}
    i\partial_t p_{\mu, s}^{I} = U^I p_{\mu, s}^{I} - \frac{1}{Z} \sum_{\nu, X} T_{\nu\mu} \langle \hat{N}_{\mu, \bar{s}}^I \rangle^0 p_{\nu, s}^{X}, \label{eq:factorised}
\end{equation}

\noindent
where to our order of approximation, 
$\langle \hat{N}_{\mu, \bar{s}}^I \rangle \approx \langle \hat{N}_{\mu, \bar{s}}^I \rangle^0$ 
is sufficiently approximated to order $\mathcal{O}(1)$. 
A similar equation could be found using other approximation schemes, 
e.g. \cite{fischer2008bogoliubov}.

Solving the above, we obtain holon (dominated by the $p_{\mu,s}^0$ contributions) 
and doublon (dominated by the $p_{\mu, s}^1$ contributions, with energies centred around $U$) 
eigen-states and eigen-energies. 
Note that the holon energies are actually minus the corresponding eigen-energies.

In order to explicitly study the effects of different lattice geometries, 
we keep $T/Z$ constant between lattices.

\begin{figure*}
    \centering
    \includegraphics[width=\textwidth]{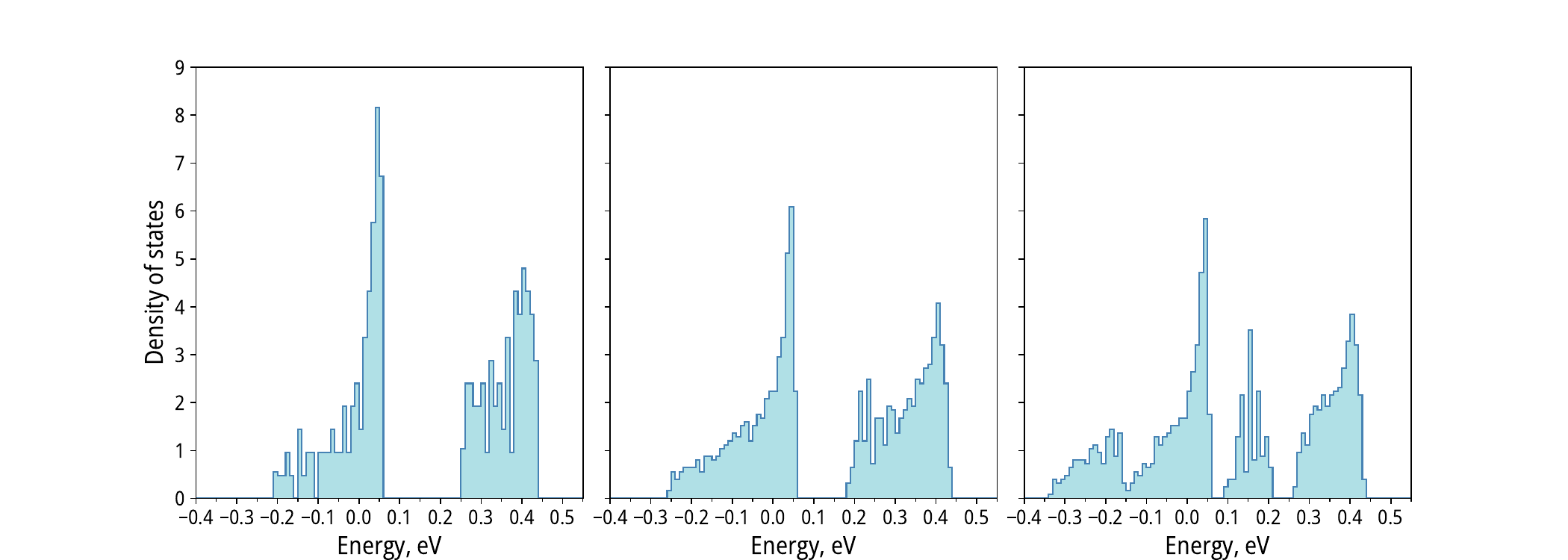}
    \caption{Holon and doublon quasi-particle excitation eigen-energy spectra obtained using the hierarchy of correlations for the hexagonal lattice with charge disorder at $r=26\%$ for (a) $V=0.00$ eV (b) $V=-0.10$ eV (c) $V=-0.20$ eV.}
    \label{fig:hc_charge_disorder_spectrum}
\end{figure*}

\subsection{Charge disorder \label{sec:hc_charge_disorder}}

We consider 
a system at half-filling in the large-$U$ limit. 
This imposes that every lattice site has one fermion, but the spin direction is not fixed. 
Thus, as a starting point, we assume 
the same statistic mixture of the two spin states at each lattice site 

\begin{equation}
    \hat{\rho}_{\mu}^0 = \frac{1}{2}\left( \ket{\uparrow}_{\mu}\bra{\uparrow} + \ket{\downarrow}_{\mu}\bra{\downarrow} \right).
\end{equation}

This immediately gives that $\langle \hat{N}_{\mu, \bar{s}}^I \rangle^0 = 1/2$ 
in equation \eqref{eq:factorised}. 
Adding the charge disorder of equation \eqref{eq:charge_disorder_hamiltonian}, 
quasi-particles are now described by

\begin{equation}
    i\partial_t p_{\mu, s}^{I} = \left(U^I + V_{\mu} \right) p_{\mu, s}^{I} - \frac{1}{2Z} \sum_{\nu, X} T_{\nu\mu} p_{\nu, s}^{X} \label{eq:factorised_charge_disorder},
\end{equation}
which we solve numerically.

In the following, 
we fix $U = 0.35$ eV and $T / Z = 0.05$~eV.
Eigen-spectra for $V\in \{0.00,-0.10,-0.20\}$~eV for identical charge distributions 
are shown in Figure \ref{fig:hc_charge_disorder_spectrum}. 
Note that we observe the splitting of holon and doublon bands into sub-bands, 
separated in energy by the on-site potential strength $V$.

We can take a step further and also characterise the eigen-states by how localised they are. 
A simple quantifier for this is the inverse participation ratio (IPR), 
defined in our case by the functional \cite{bell1970atomic, wegner1980inverse}

\begin{equation}
    \mathrm{IPR}\{p(E_n)\} = \sum_{\mu, I} | p_{\mu}^{I}(E_n) |^4.
\end{equation}

\begin{figure}
    \centering
    \includegraphics[width=0.43\textwidth]{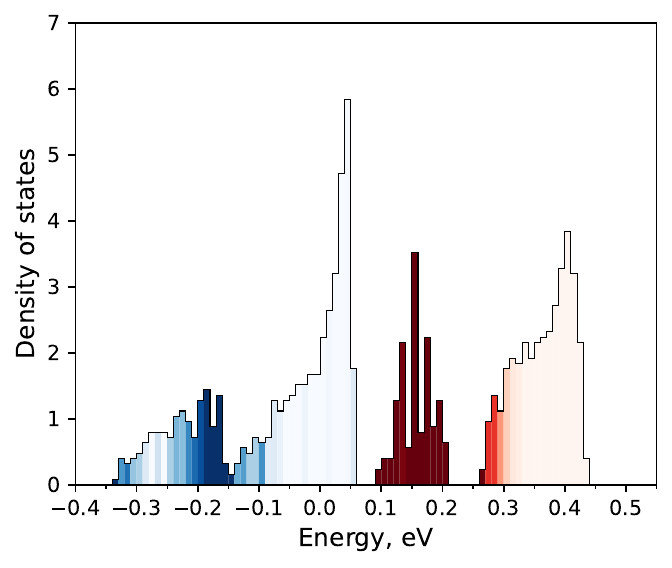}
    \caption{Holon (blue) and doublon (red) eigen-energy spectra for charge disordered hexagonal lattice with $V=-0.20$~eV and $r=26\%$. The shading visualises the proportion of localised states, with darker colours corresponding to more localised states. 
    }
    \label{fig:hc_charge_disorder_holon_doublon_localisation_colored}
\end{figure}

The inverse of the IPR, the participation ratio, describes the number of sites over which 
a state extends \cite{bell1970atomic}. 
Thus, for our normalised `eigen-states' this is bounded by $1 < \mathrm{IPR}^{-1} < 2N^2$. 
We can use this to colour the spectrum by classifying states into localised and delocalised 
states depending on their participation ratio i.e. `number of sites they cover on the lattice'. 
From e.g. Figure \ref{fig:hc_charge_disorder_holon_doublon_localisation_colored}, 
we see that the emerging sub-bands are predominantly composed of localised states, 
while the remains of the original bands are predominantly delocalised. 
Intuitively, in the limit where only a small fraction of sites are affected by the 
charge disorder, only a small number of the free quasi-particle states will be pinned 
down by the disorder, the rest remaining as free, plane-wave-like states.

\begin{figure*}
    \centering
    \subfigure[]{\includegraphics[width=0.3\textwidth]{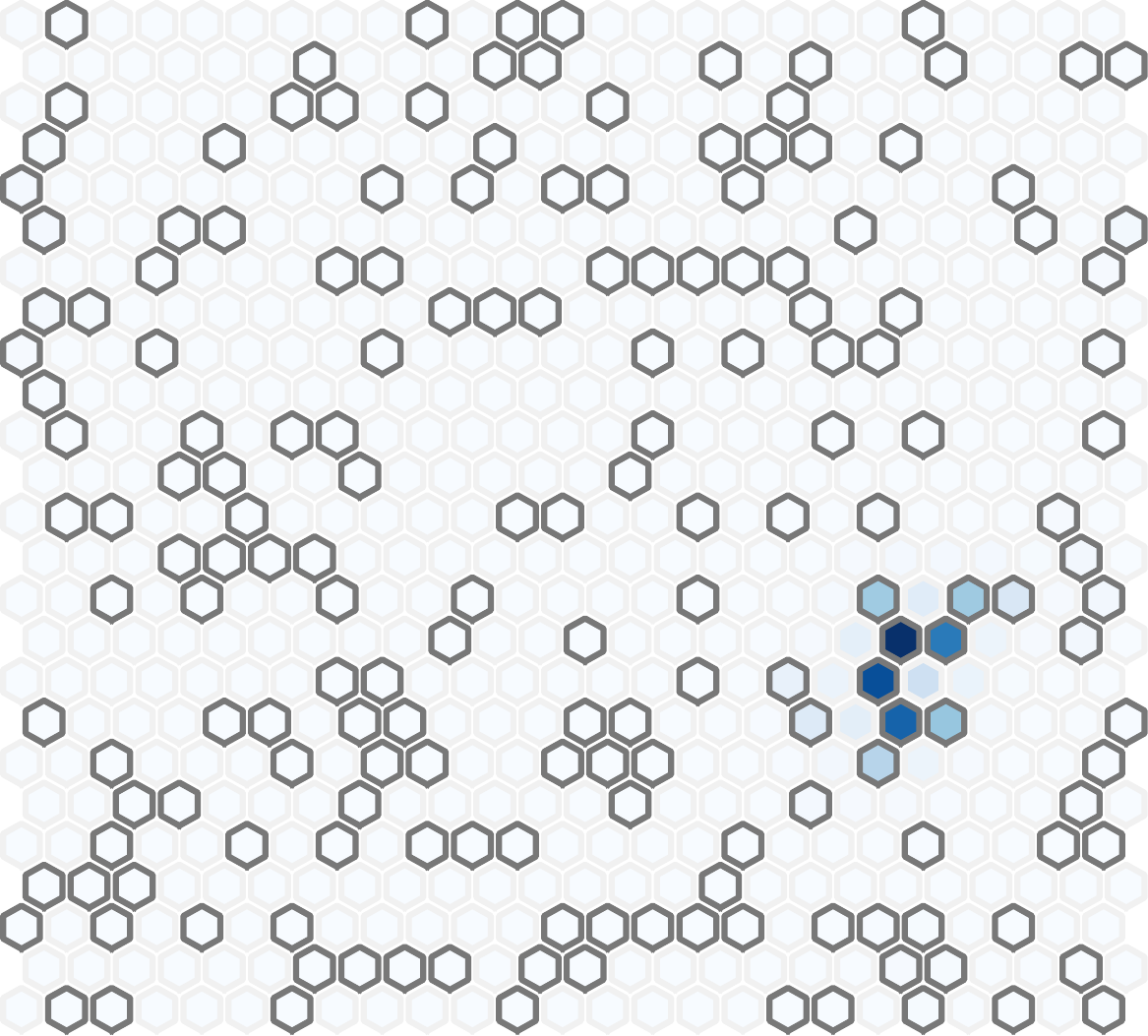}}
    \hspace{0.4cm}
    \subfigure[]{\includegraphics[width=0.3\textwidth]{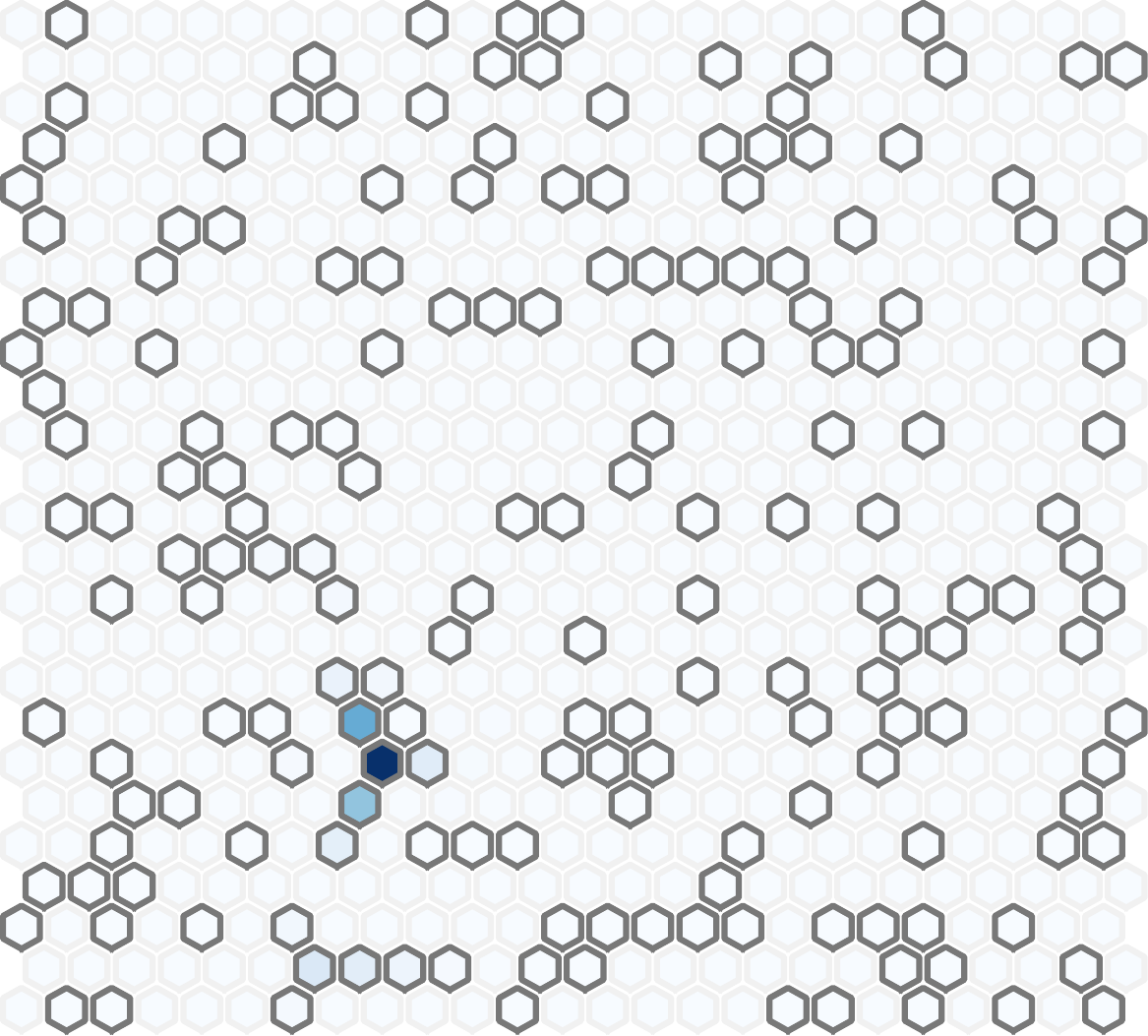}}
    \hspace{0.4cm}
    \subfigure[]{\includegraphics[width=0.3\textwidth]{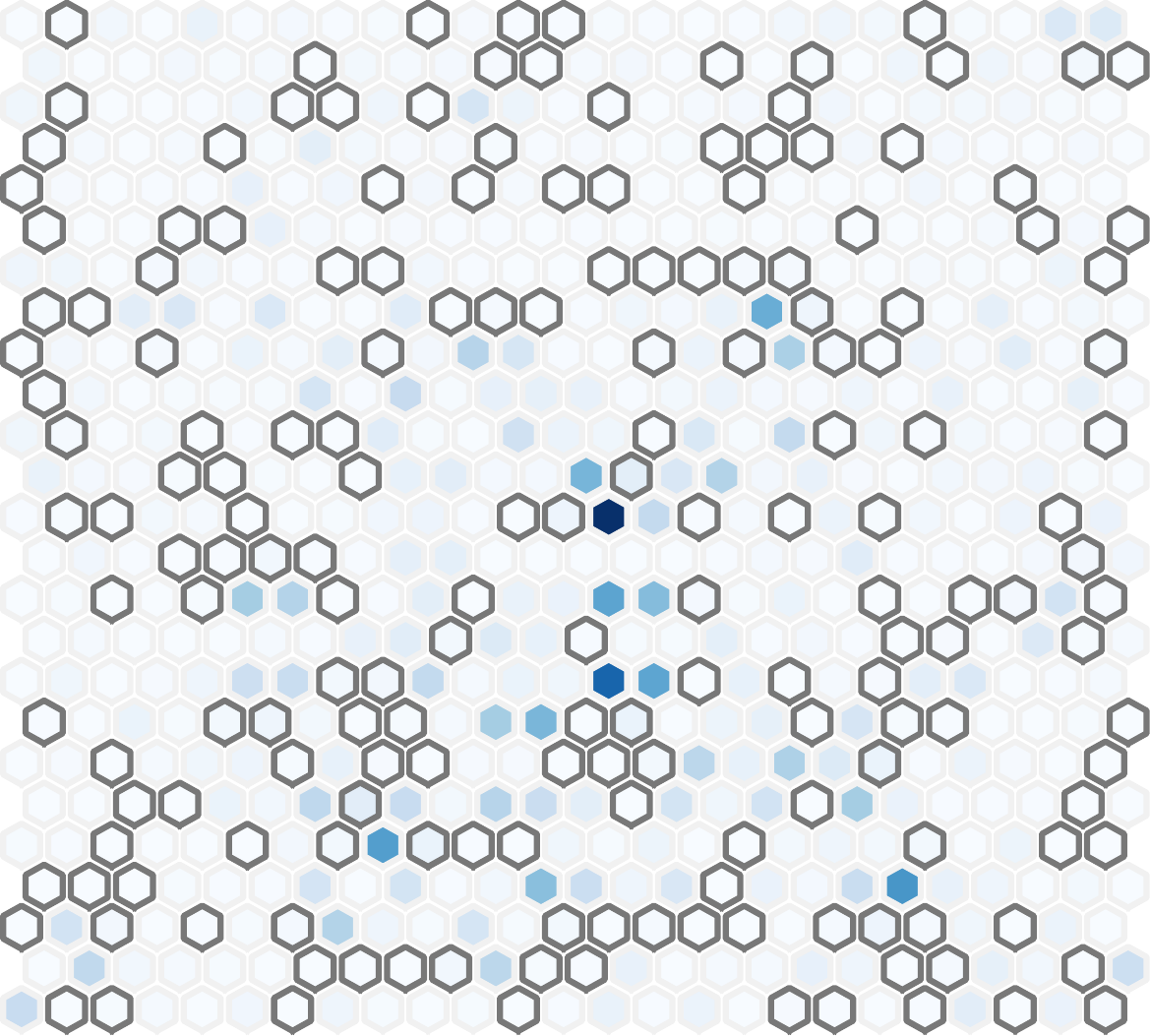}}
    \caption{Illustrations of selected holon eigen-states in the charge disordered hexagonal lattice. Dark outline corresponds to sites with on-site energy $V_{\mu} = V=-0.20$~eV, light outline corresponds to $V_{\mu} = 0$. The states have eigen-energies (a) $E\approx -0.315$ eV (bottom of doped holon sub-band) (b) $E\approx -0.161$ eV (top of doped holon sub-band) (c) $E\approx -0.031$ eV (middle of undoped holon sub-band).}
    \label{fig:hc_charge_disorder_holon_eigenstates}
\end{figure*}

\begin{figure}
    \centering
    \includegraphics[width=0.43\textwidth]{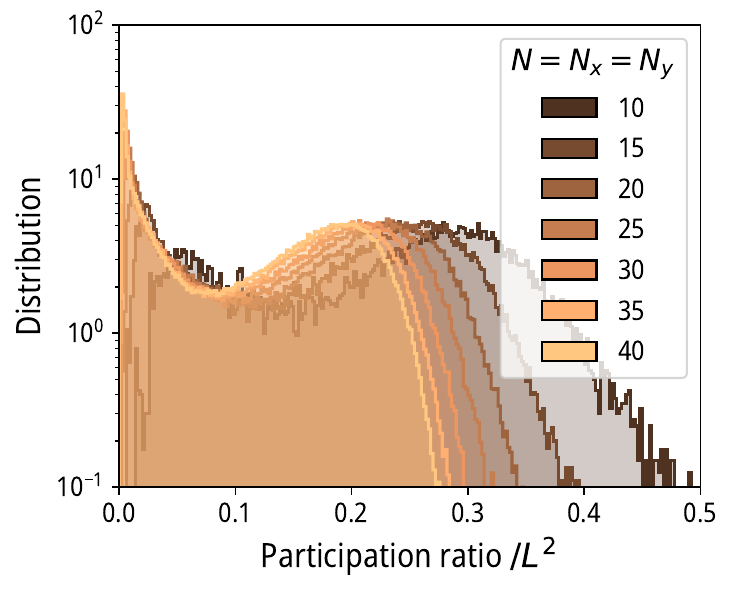}
    \caption{Normalised distributions of the participation ratio for different lattice sizes in the case of charge disorder characterised by $V=-0.20$ eV and $r=26\%$. 
    \label{fig:hc_charge_disorder_size_change_localisation_distribution}}
\end{figure}

Selected holon states are visualised in Figure \ref{fig:hc_charge_disorder_holon_eigenstates} 
for different eigen-energies in the holon band. 
We can go a step further and plot the scaling of the participation ratio distribution as the 
lattice size changes - this is shown in Figure 
\ref{fig:hc_charge_disorder_size_change_localisation_distribution}. 
Interestingly, we observe two distinct peaks in the distribution, one corresponding to 
very localised states, the second corresponding to delocalised states. 
Thus, the charge disorder directly results in the emergence of two different length 
scales in the system, corresponding to the typical sizes of localised and delocalised states.

Finally, note that the fine structure (peaks) in the localised doublon sub-band in 
Figure \ref{fig:hc_charge_disorder_holon_doublon_localisation_colored} persists also 
after averaging the spectrum over many disorder realisations. 
Physically, the central peak can be considered to be due to single, isolated doped sites, 
while the surrounding two peaks arise as the symmetric and anti-symmetric states on 
small clusters of doped sites.


\subsubsection{Square and honeycomb lattices}

\begin{figure*}
    \centering
    \subfigure[]{
    \includegraphics[width=0.43\textwidth]{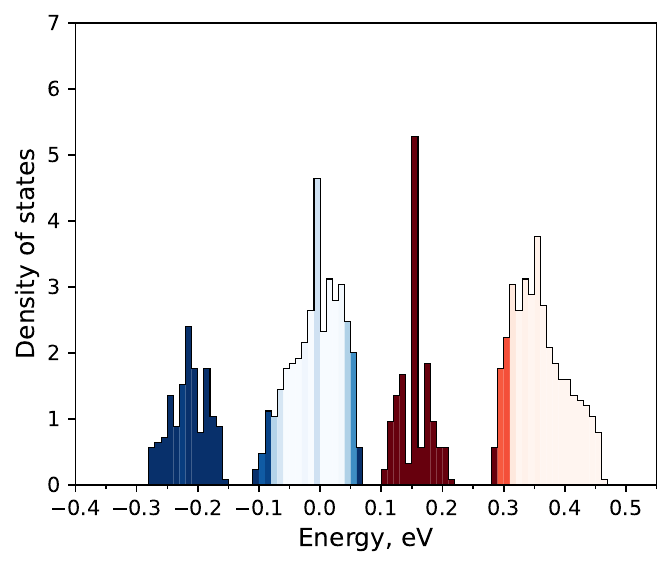}
    \label{fig:hc_charge_disorder_holon_doublon_localisation_colored_square}}
    \subfigure[]{
    \includegraphics[width=0.43\textwidth]{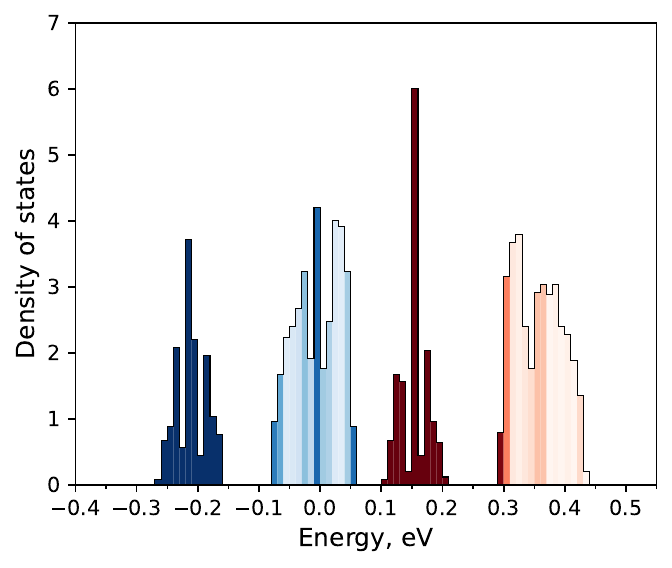}
    \label{fig:hc_charge_disorder_holon_doublon_localisation_colored_honeycomb}}
    \caption{Holon (blue) and doublon (red) eigen-energy spectra for charge disordered (a) square and (b) honeycomb lattices with $V=-0.20$ eV and $r=26\%$. The shading visualises the proportion of localised states, with darker colours corresponding to more localised states. 
    }
\end{figure*}

\begin{figure}
    \centering
    \includegraphics[width=0.43\textwidth]{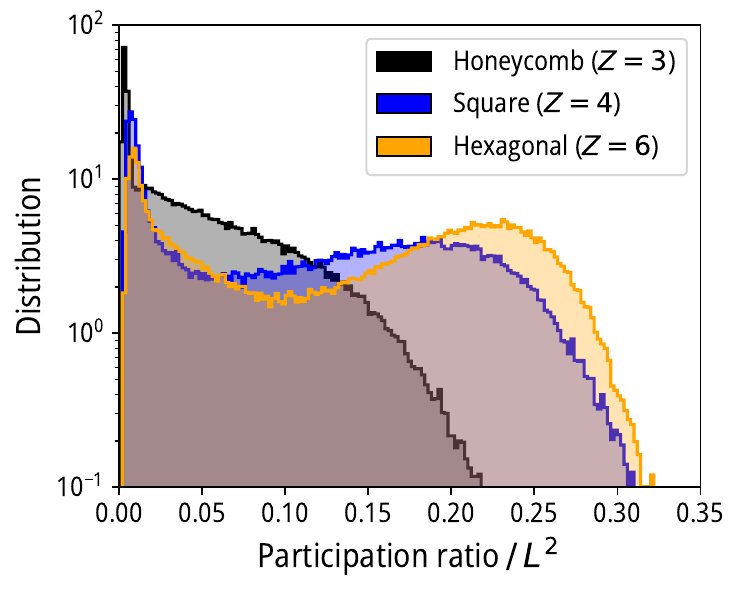}
    \caption{Normalised distributions of the participation ratio in hexagonal, square, and honeycomb lattices with charge disorder characterised by $V=-0.20$ eV and $r=26\%$. 
    }
    \label{fig:hc_charge_disorder_all_lattices_localisation_density}
\end{figure}

Comparison between hexagonal, 
square and honeycomb lattices offers insight into how reducing the coordination number 
$Z$ affects the quasi-particle states. 
From a site percolation theory perspective, with the same fraction of doped sites, 
having fewer neighbours increases the threshold below which finite clusters dominate 
the physics of the system \cite{essam1964exact}. 
Thus, for instance, there is a noticeable difference in fraction of very localised sites 
between the hexagonal lattice in Figure 
\ref{fig:hc_charge_disorder_holon_doublon_localisation_colored} 
and the square and honeycomb lattices in Figures 
\ref{fig:hc_charge_disorder_holon_doublon_localisation_colored_square} 
and \ref{fig:hc_charge_disorder_holon_doublon_localisation_colored_honeycomb}, respectively. 
This becomes even more apparent if we consider the distribution of participation ratios, 
Figure \ref{fig:hc_charge_disorder_all_lattices_localisation_density}. 
Note that the bimodal structure observed in the previous subsection exists for all three 
lattice types, although the separation of the two length scales appears to be strongly 
dependent on the lattice type.

\subsection{Spin-fixed background}

Consider now a half-filled background where each site contains one electron with a fixed spin direction. Mathematically, this means

\begin{equation}
    \hat{\rho}_{\mu}^0 \in \{ \ket{\uparrow}_{\mu}\bra{\uparrow}, \ket{\downarrow}_{\mu}\bra{\downarrow} \},
\end{equation}

\noindent
and correspondingly $\langle \hat{N}_{\mu, \bar{s}}^I \rangle^0 \in \{0, 1\}$ depending 
on the orientation of the spin at site $\mu$. 
We maintain a spin-unpolarised sample, meaning $\hat{\rho}_{\mu}^0$ has equal probability 
to be either of the two options above.

Consider now the factorised equation of motion \eqref{eq:factorised}. 
If $\langle \hat{N}_{\mu, \bar{s}}^I \rangle^0 = 0$, then we simply get diagonal 
evolution of $p_{\mu, s}^{I}$ without any coupling to adjacent sites. 
For example, consider a site $\mu$ which has an electron of spin $s$ in the background. 
Then, $\langle \hat{N}_{\mu, \bar{s}}^1 \rangle^0 = \langle \hat{N}_{\mu, s}^0 \rangle^0 = 0$ 
and $\langle \hat{N}_{\mu, \bar{s}}^0 \rangle^0 = \langle \hat{N}_{\mu, s}^1 \rangle^0 = 1$. 
Thus, here the diagonal terms are $p_{\mu, s}^{1}$ and $p_{\mu, \bar{s}}^{0}$. 
We now argue that such terms should, in fact, be discarded from our consideration.

First, recall that $p_{\mu, s}^I$ is defined through the factorisation in 
equation \eqref{eq:factorisation_ansatz}. 
Considering now arbitrary $\hat{C}_{\nu, s}^{J \dagger}$ at $t=0$,

\begin{equation}
    s: \langle \hat{N}_{\mu, s}^1 \rangle^0 = 1 \implies \left( p_{\nu, s}^J \right)^* p_{\mu, s}^1 = \langle \hat{C}_{\nu, s}^{J \dagger} \hat{C}_{\mu, s}^{1} \rangle = 0,
\end{equation}

\noindent
and hence $p_{\mu, s}^1(t=0) = 0$. 
A similar consideration also fixes $p_{\mu, \bar{s}}^0(t=0) = 0$. 
Then, because $p_{\mu, s}^{1}$ and $p_{\mu, \bar{s}}^{0}$ evolve diagonally, 
they also stay zero at all times to this order of approximation. 
Physically, this is the statement that the evolution of the background is not taken 
into account when considering the two-point correlators to leading order in $1/Z$. 
Note that these effects can become important at higher orders of the hierarchy of 
correlations \cite{queisser2024back}.

Thus, only one in each pair $\{p_{\mu, s}^{I}, p_{\mu, \bar{s}}^{I}\}$ is relevant, 
and in what follows we can discard the other from our considerations. 
Correspondingly, for the remainder of this section we omit the spin index, 
since this is implicitly fixed by the positional index $\mu$ and superscript $I$.

Finally, the relevant factorised equations for the spin-fixed background read

\begin{equation}
    i\partial_t p_{\mu}^{I} = U^I p_{\mu}^{I} - \frac{1}{Z} \sum_{\nu} T_{\nu\mu} \left( p_{\nu}^{I} \delta_{\sigma_{\mu},\sigma_{\nu}} + p_{\nu}^{1-I} \delta_{\sigma_{\mu},\bar{\sigma}_{\nu}} \right),
\end{equation}

\noindent
where $\sigma_{\mu}$ is the spin at site $\mu$. 
Note that, by throwing away the irrelevant $p$'s, we are immediately reassured that 
the time evolution of $p_{\mu, s}^I$ is unitary -- something that we could not necessarily 
have concluded otherwise. 
Note also that the two relevant $p_{\mu, s}^I$ at each site are completely decoupled.

\begin{figure}
    \centering
    \includegraphics[width=0.43\textwidth]{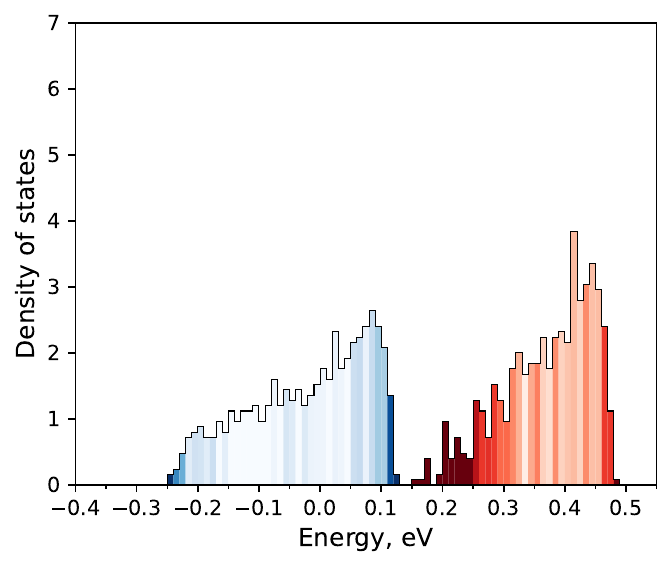}
    \caption{Holon (blue) and doublon (red) energy spectra for spin disordered hexagonal lattice. The shading visualises the proportion of localised states, with darker colour corresponding to more localised states. 
    }
    \label{fig:hc_spin_disorder_holon_doublon_localisation_colored}
\end{figure}

Note that even in this case the spectrum shows clearly distinct holon and doublon bands, 
see Figure \ref{fig:hc_spin_disorder_holon_doublon_localisation_colored}. 
On the other hand, in contrast to the charge disorder case, there is no splitting of 
the bands into sub-bands, while localised states appear to be present at all energies 
within the bands, not just the band edges.

\subsection{Combined disorder}

\begin{figure}
    \centering
    \includegraphics[width=0.43\textwidth]{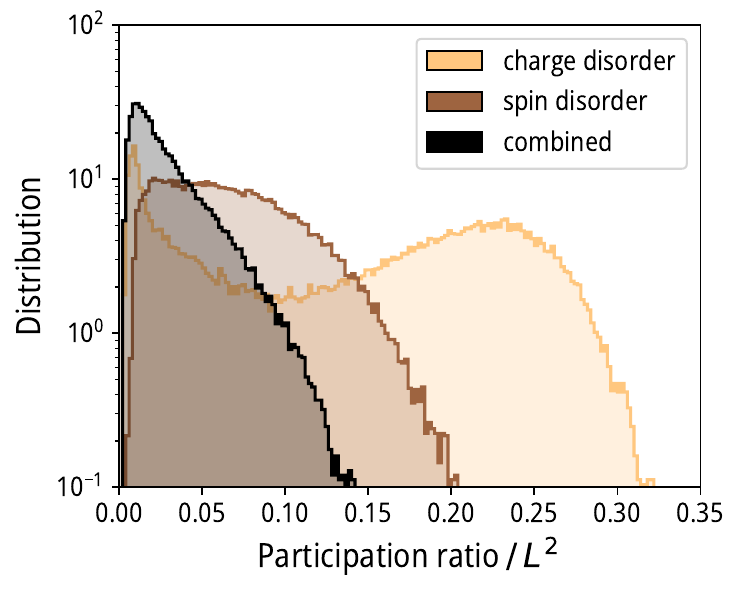}
    \caption{Density of states versus the participation ratio for spin, charge and a combination of disorders for the hexagonal lattice. Charge disorder with parameters $V=-0.20$~eV and $r=26\%$. 
    }
    \label{fig:hc_combined_disorder_localisation_density}
\end{figure}

Including both the spin disorder and charge disorder, we expect even stronger localisation 
of the eigen-states. 
This is indeed what we observe, and is most clearly seen by comparing the distributions 
of participation ratios for the three different cases of charge disorder, spin disorder, 
and combined disorder, see Figure \ref{fig:hc_combined_disorder_localisation_density}.

\section{Perturbation theory in $T/U$}

It is instructive to compare results obtained using the hierarchy of correlations with 
a more conventional method. 
To this end, we attempt to reproduce the most important points of the previous section, 
now from the perspective of strong-coupling perturbation theory. 
We focus on holon excitations on the half-filled background, and, to make the 
calculations tractable, restrict ourselves to first order. 
Further, we restrict ourselves to the single-hole subspace of the full Hilbert space, 
meaning that we do not consider states with e.g. two holes and one doubly-occupied site. 
Including these would affect the energy spectrum at second order. 
However, this cannot be said for the eigen-states, since already for $V=-0.20$ eV we could 
have the energy of a state with a doubly occupied site plus two hole sites be comparable 
to the energy of a single hole state, $U+V \sim V$.

We work in the regime of small hopping parameter $T$. 
Consequently, our starting point in the perturbation theory are single-site states.

\subsection{Charge disorder}


Formally, a perturbation theory approach to a spin-disordered background would 
necessitate considering the entire space of states differing by permutations of the spins. 
However, due to the arbitrary charge disorder, this is intractable analytically and 
numerically for all but the smallest lattices.

As a simplified picture, we approximate the relevant dynamics by
\begin{enumerate}
    \item imposing as a constraint on the Hilbert space that each lattice site can 
    contain at most one fermion;
    \item taking $T \rightarrow T/2$ to capture the effectively suppressed hopping 
    in a disordered background.
\end{enumerate}

As discussed, the first of these already fails when considering first order corrections 
to eigen-states of the system. 
The second is an \textit{ad hoc} way to reproduce the correct energy spectrum in a pristine 
lattice. 
We could also motivate this by similarity with section \ref{sec:hc_charge_disorder}.

\begin{figure*}
    \centering
    \includegraphics[width=\textwidth]{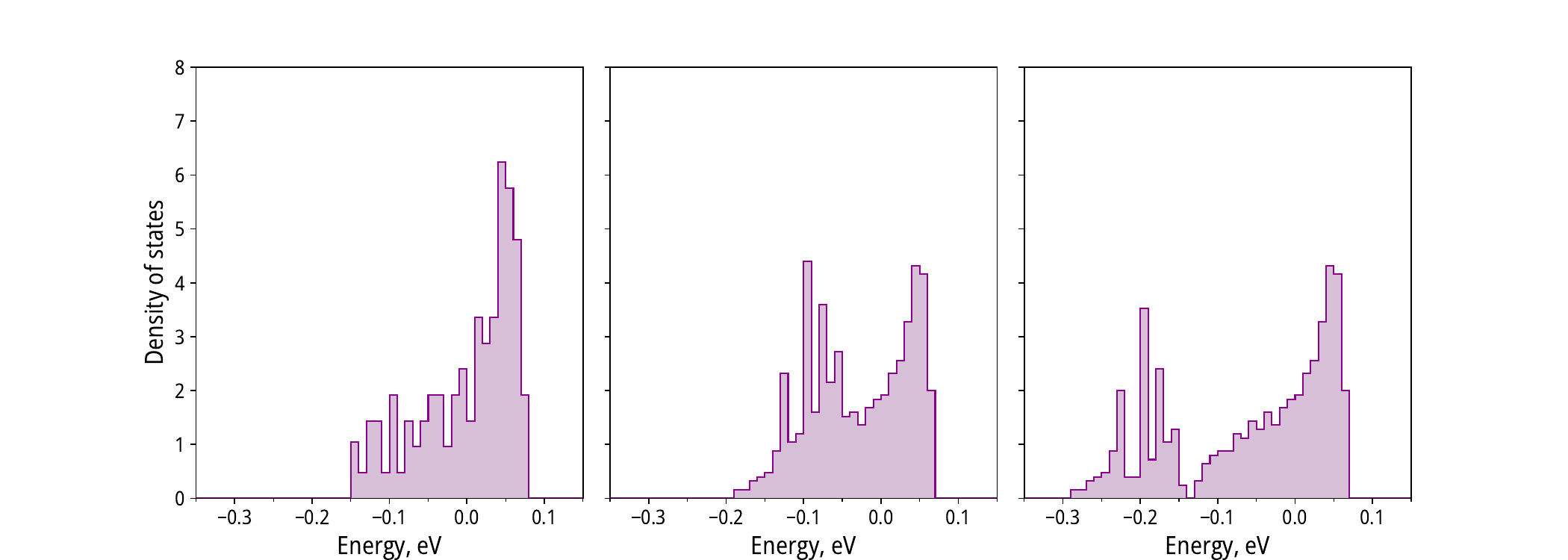}
    \caption{Holon eigen-energy spectra obtained using perturbation theory in the hexagonal lattice with charge disorder at $r=26\%$ for (a) $V=0.00$ eV (b) $V=-0.10$ eV (c) $V=-0.20$ eV. 
    To facilitate comparison with Figure \ref{fig:hc_charge_disorder_spectrum}, 
    the density of states is divided by 2.}
    \label{fig:pt_charge_disorder_spectrum}
\end{figure*}

\begin{figure}
    \centering
    \includegraphics[width=0.43\textwidth]{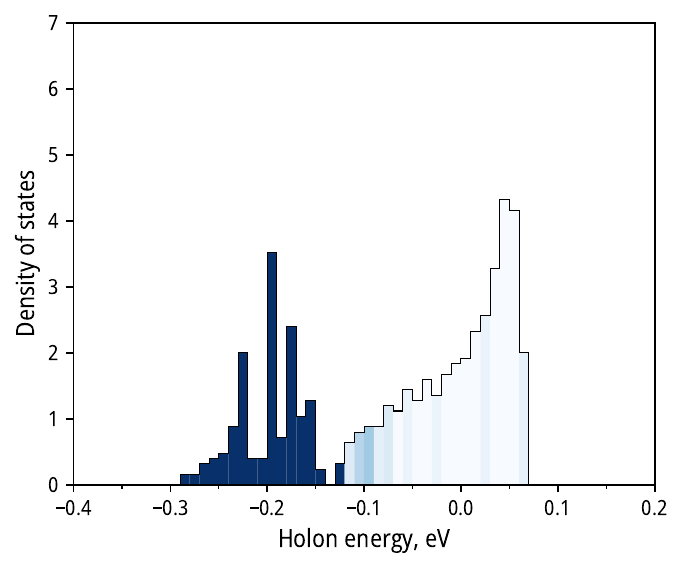}
    \caption{Holon eigen-energy spectrum for charge disordered hexagonal lattice, coloured to reflect the localisation of states in each energy range. Darker shading corresponds to a larger proportion of localised states.  
    To facilitate comparison with Figure \ref{fig:hc_charge_disorder_holon_doublon_localisation_colored}, the density of states is divided by 2.}
    \label{fig:pt_charge_disorder_holon_localisation_colored}
\end{figure}

The eigen-spectra, with sign of energy inverted to facilitate comparison with previous results, 
are shown in Figure \ref{fig:pt_charge_disorder_spectrum} for $V\in \{0.00,-0.10,-0.20\}$~eV. 
Qualitatively, these capture the same splitting into two sub-bands observed already in 
Figure \ref{fig:hc_charge_disorder_spectrum}. 
In fact, considering also the colored spectrum for $V=-0.20$ eV in 
Figure \ref{fig:pt_charge_disorder_holon_localisation_colored}, the structure of the 
localised sub-band is actually closer to that of the localised doublon sub-band obtained 
using the hierarchy of correlations. 
We expect this fine structure to be smoothed out at higher orders in perturbation theory 
due to mixing with states not included in our restricted Hilbert space at first order. 
Thus, the hierarchy of correlations at first order in $1/Z$ already captures at least some 
of the effects that only appear beyond first order in perturbation theory.

\subsection{Spin-fixed background}

\begin{figure}
    \centering
    \includegraphics[width=0.43\textwidth]{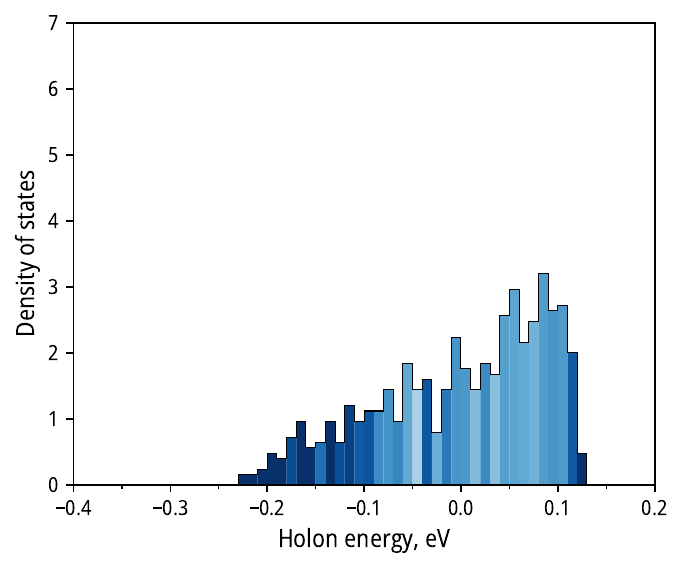}
    \caption{Holon eigen-energy spectrum for spin disordered hexagonal lattice, coloured to reflect the localisation of states in each energy range. Darker shading corresponds to a larger proportion of localised states.  
    To facilitate comparison with Figure \ref{fig:hc_spin_disorder_holon_doublon_localisation_colored}, the density of states is divided by 2.}
    \label{fig:pt_spin_disorder_holon_localisation_colored}
\end{figure}

Consider now the half-filled background where each site contains a single fermion, 
its direction fixed in the ground state by the spin splitting Hamiltonian of 
equation \eqref{eq:spin_splitting_hamiltonian}. 
Figure \ref{fig:pt_spin_disorder_holon_localisation_colored} shows that perturbation 
theory predicts the states in the holon band to be more localised than suggested 
by the analogous results from the hierarchy of correlations, 
Figure \ref{fig:hc_spin_disorder_holon_doublon_localisation_colored}. 
Following the discussion of the previous subsection, it can again be suggested that 
this is due to the hierarchy of correlations at first order in $1/Z$ capturing 
interactions between more states, which we have excluded from our perturbation 
theory analysis to make calculations in the latter tractable.

\section{Conclusions and Outlook}

Through an analysis based on the hierarchy of correlations, as well as a comparison 
with strong-coupling perturbation theory, we have analysed quasi-particle states in 
the half-filled Fermi-Hubbard model in the presence of charge and spin disorders. 
Our results show the states localising to varying degrees. 

Adding spin disorder to the system restricts the types of excitations that can be present 
at each lattice site. 
The quasi-particle states localise to varying degrees, depending on how well each 
quasi-particle fits into the new, restricted lattice space. 
Consequently, localisation occurs at energies throughout the original quasi-particle bands.

An intriguing contrast to this is provided by charge disorder, which results in a distinct 
bimodal distribution of the participation ratios of eigen-states, indicating the emergence 
of two distinct length scales corresponding to localised and delocalised states. 
Furthermore, the localised states are observed to form localised holon and doublon sub-bands 
in the energy eigen-spectra. These separate energetically from what remains of the 
(delocalised) original quasi-particle bands, with the separation controlled by the 
on-site disorder potential strength $V$. 
In turn, this also indicates a spatial separation between localised states in doped regions 
(at potential $V$) and undoped regions (at zero potential). 


\section*{Acknowledgments}

This work was funded by the Deutsche Forschungsgemeinschaft (DFG, German Research Foundation) through Project No. 278162697 (SFB 1242).

\interlinepenalty=10000
\bibliography{library.bib}
    
\end{document}